\pdfoutput=1
\documentclass[aps,prl,showpacs,reprint]{revtex4-1}
\usepackage{graphicx}
\usepackage{subfigure}
\usepackage{amsmath}
\usepackage{amssymb}
\usepackage{natbib}

\usepackage{esvect}

\usepackage{color} 

\usepackage{hyperref}
\hypersetup{
colorlinks=true,final=true,
        linkcolor=blue,
        citecolor=blue,
        filecolor=blue,
        urlcolor=blue,
}

\begin{document}

\title{All-Microwave Control and Dispersive Readout of Gate-Defined Quantum Dot Qubits \\
in Circuit Quantum Electrodynamics
}
\author{P.~Scarlino}
\thanks{These authors contributed equally to this work.}
\author{D.~J.~van~Woerkom}
\thanks{These authors contributed equally to this work.}
\author{A.~Stockklauser}
\author{J.~V.~Koski}
\author{M.~C.~Collodo}
\author{S.~Gasparinetti}
\author{C.~Reichl}
\author{W.~Wegscheider}
\author{T.~Ihn}
\author{K.~Ensslin}
\author{A.~Wallraff}
\affiliation{Department of Physics, ETH Zurich, CH-8093 Zurich, Switzerland}
\date{\today}
\begin{abstract}

Developing fast and accurate control and readout techniques is an important challenge in quantum information processing with semiconductor qubits. Here, we study the dynamics and the coherence properties of a GaAs/AlGaAs double quantum dot (DQD) charge qubit strongly coupled to a high-impedance SQUID array resonator. We drive qubit transitions with synthesized microwave pulses and perform qubit readout through the state dependent frequency shift imparted by the qubit on the dispersively coupled resonator. We perform Rabi oscillation, Ramsey fringe, energy relaxation and Hahn-echo measurements and find significantly reduced decoherence rates down to $\gamma_2/2\pi\sim 3\,\rm{MHz}$ corresponding to coherence times of up to $T_2 \sim 50 \, \rm{ns}$ for charge states in gate defined quantum dot qubits.
\end{abstract}
\maketitle

Fundamental and applied research on semiconductor quantum dots \cite{Loss1998,kouwenhoven2001,Wiel2002} attracts much attention largely due to the potential of using the electron charge \cite{Hayashi2003,Petta2004} and spin \cite{Petta2005,Hanson2007} degrees of freedom as information carriers in solid state qubits. In practice, the coherence of both spin and charge qubits are limited by charge noise \cite{Hayashi2003, Petersson2010, dial2013, Paladino2014, reed2016, martins2016}. As a consequence, both improving coherence properties and reducing the time scale for control and readout of qubits are important topics of current research as they are crucial for realizing quantum information processing in such systems.

In this work, we address both challenges by making use of strong coherent coupling between charges in double quantum dots and photons stored in an on-chip resonator using the circuit QED architecture realized first with superconducting qubits \cite{wallraff2004} and more recently in Silicon \cite{Mi2017} and GaAs \cite{Stockklauser2017} quantum nanostructures. We use non-resonant (dispersive) interactions between a DQD charge qubit and a high impedance resonator for time-resolved readout of the qubit coherently manipulated using microwave pulses. We employ this technique to investigate coherence properties of qubits.

In a more conventional approach, in which DQD charge qubits are manipulated using non-adiabatic pulses and read out by capacitively coupled charge detectors, $T_2$ coherence times of up to $(7 \pm 2.5)\,\rm{ns}$ in GaAs \cite{Petersson2010} and $(2.1 \pm 0.4)\,\rm{ns}$ in Si \cite{shi2013c} have been observed. Spin Echo experiments performed with non-adiabatic pulses \cite{dovzhenko2011} found echo times of $T_\mathrm{2,echo} \sim (1.4 \pm 0.4)\,\rm{ns}$ in GaAs \cite{wang2017g} and Si \cite{shi2013c}. Recently, first microwave driven coherent operations showed improved qubit control and a $T_\mathrm{2,echo} \sim (2.2 \pm 0.1)\,\rm{ns}$ in three electron DQDs in SiGe \cite{Kim2015a}, which is operated as hybrid spin and charge qubit. In a more recent work, the free induction decay time of this system has been extended to $T_2\sim 177\,\rm{ns}$ through operation in the spin-like operating region \cite{thorgrimsson2017}.

\begin{figure}[!b]
\includegraphics[width=\columnwidth]{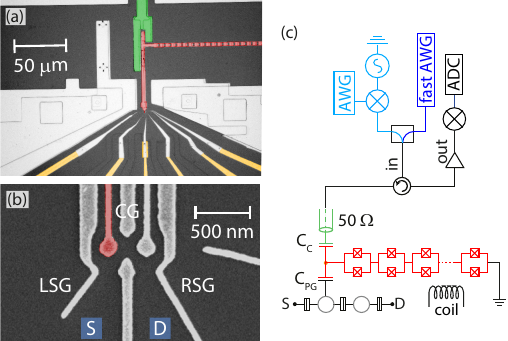}
\caption{Sample and simplified circuit diagram. Optical micrograph of the device showing the substrate (dark gray), the Au gate leads (yellow), the superconducting structures including the Al fine gate structure forming the DQD (light gray), the SQUID array (red) and the microwave feedline (green). (b) SEM micrograph of the DQD showing its superconducting Al top gates (gray) and the plunger gate coupled to the resonator (red). (c) Circuit diagram schematically displaying the DQD source contact ($S$), drain contact ($D$), and coupling capacitance to the resonator ($C_{\rm{PG}}$) and essential components in the microwave detection chain (circulator, amplifier) used for performing reflectance measurements on the device. Boxes with crosses and rectangles indicate Josephson and normal tunnel junctions, respectively.
}
\label{fig:SampleAndCircuit}
\end{figure}

Here, we perform experiments with a superconducting high-impedance resonator coupled to a DQD charge qubit \cite{Stockklauser2017}. The high-impedance resonator, which boosts the coupling strength and is frequency tunable, is composed of an array of $32$ SQUIDs [Fig.~\ref{fig:SampleAndCircuit}(a)] grounded at one end and terminated at the other end by a small island that is capacitively coupled to a coplanar waveguide used as a drive line. A gate line [red in Fig.~\ref{fig:SampleAndCircuit}(b)] extends from the island towards the DQD forming one of its plunger gates. The DQD is defined using voltage-biased Aluminum (Al) depletion gates \cite{Wiel2002} connected to gold (Au) leads deposited on a small mesa etched into a GaAs/AlGaAs heterostructure forming a two dimensional electron gas (2DEG) 90 nm below the surface [Figs.~\ref{fig:SampleAndCircuit}(a) and (b)]. We estimate an electron number of around 10 in each QD from the respective charging energies ($E_{\rm{c},1} \sim E_{\rm{c},2}  \sim E_{\rm{c},m}  \sim 100 \,\rm{GHz}$) \cite{Wiel2002}. The device is operated in a dilution refrigerator at a temperature of $\sim 30 \,\rm{mK}$.

\begin{figure}[!b]
\includegraphics[width=\columnwidth]{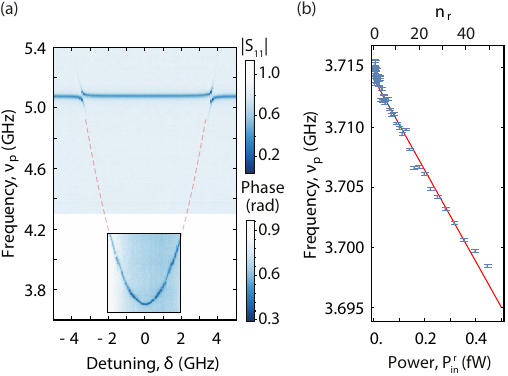}
\caption{Resonator and two-tone DQD charge qubit spectroscopy.
(a) Resonator reflection response $|S_{11}|$ and qubit spectroscopy (inset: phase response of microwave tone applied at the resonator frequency) versus DQD detuning $\delta$, with $2t=3.71\,\rm{GHz}$, $\nu_{\rm{r}}(\Phi \sim 0.3 \Phi_0) =5.070\,\rm{GHz}$ and $2g/2\pi=75\,\rm{MHz}$.
(b) AC-Stark shift of the DQD at $\delta=0$ as a function of the power $P_{\rm{in}}^{\rm{r}}$ at the resonator input, applied at the resonator frequency $\nu_\mathrm{p}=\nu_\mathrm{r}=5.070 \,\mathrm{GHz}$.
Measured DQD qubit frequency $\tilde{\omega}_\mathrm{q}=\omega_\mathrm{q}(P_{\rm{s}} \rightarrow 0)+2n_{\rm{r}}g^2/\Delta_\mathrm{r,q}$ and fit (solid line) \textsl{vs.}~$P_{\rm{in}}^{\rm{r}}$. The intra-resonator photon number $n_{\rm{r}}$ extracted from the fit is indicated on the top axis.}
\label{fig:ResDQDChar}
\end{figure}

We control the DQD qubit transition frequency $\nu_{\rm{q}}=\sqrt{4t^2+\delta^2}$ by tuning the inter-dot tunnel rate $2t$ to $3.71\,\rm{GHz}$ and by adjusting the detuning $\delta$ by applying bias voltages to the respective gates \cite{Frey2012}. We tune the resonator frequency to $\nu_{\rm{r}}(\Phi \sim 0.3 \Phi_0) = 5.07\,\rm{GHz}$ using externally applied magnetic flux $\Phi$
\cite{Stockklauser2017}. Sweeping the DQD detuning $\delta$, the reflectance spectrum $|S_{11}(\nu_{\rm{p}})|$ [Fig.~\ref{fig:ResDQDChar}(a)] shows characteristic dispersive shifts in the resonator spectrum for $\nu_{\rm{q}} \gg \nu_{\rm{r}}$ and $\nu_{\rm{q}} \ll \nu_{\rm{r}}$, and avoided crossings that are the signature of strong coupling at resonance $\nu_{\rm{q}} = \nu_{\rm{r}}$ \cite{Stockklauser2017,Mi2017}.

We measure a resonator linewidth of $\kappa_{\rm{tot}}/2\pi=\kappa_{\rm{ext}}/2\pi+\kappa_{\rm{int}}/2\pi\sim 23 + 7 \,\rm{MHz} = 30\,\rm{MHz}$ with external coupling $\kappa_{\rm{ext}}$ exceeding the internal losses $\kappa_{\rm{int}}$ realizing the overcoupled regime.
From the vacuum Rabi splitting we extract a coherent coupling strength of $g/2\pi \sim 57 \,\rm{MHz}$ between the resonator and the DQD at resonance for
$\nu_{\rm{q}}(\delta=0)=2t \sim \nu_{\rm{r}} = 5.695\,\rm{GHz}$ \cite{Stockklauser2017}.

As mentioned above, when the DQD transition frequency $\omega_{\rm{q}}$ is detuned by $\Delta_{\rm{r,q}} = \omega_{\rm{q}} - \omega_{\rm{r}}\gg g$, the resonator frequency $\tilde{\omega}_{\rm{r}}=\omega_{\rm{r}} \pm  g^2/\Delta_{\rm{r,q}}$ is dispersively shifted conditioned on the qubit state \cite{Blais2004}. We infer the qubit transition frequency from the detected phase shift of the resonator reflectance, as recently demonstrated for semiconductor DQD qubits \cite{Stockklauser2017,Mi2017} and routinely used for superconducting qubits \cite{Schuster2005}. We perform continuous wave (CW) two-tone spectroscopy of the DQD charge qubit \cite{Schuster2005,wallraff2004} by probing the amplitude and phase ($\vert \Delta \phi \vert = \tan^{-1}[2g^2/(\kappa_{\rm{tot}}\Delta_{\rm{r,q}})]$) of the resonator reflectance at fixed measurement frequency $\nu_{\rm{p}} =\nu_{\rm{r}}= 5.07\,\rm{GHz}$ while applying an additional spectroscopy tone at frequency $\nu_{\rm{s}}$ through the resonator to the DQD qubit \cite{Stockklauser2017}. The spectroscopically extracted transition frequency displayed in the lower part of Fig.~\ref{fig:ResDQDChar}(a) is in good agreement with the calculated  qubit frequency $\nu_{\rm{q}}=\sqrt{4t^2+\delta^2}$ (red dashed line) for $2t = 3.71\,\rm{GHz}$.

The dispersive coupling harnessed for qubit readout also leads to a qubit frequency shift $2n_{\rm{r}} g^2/\Delta_{\rm{r,q}}$, known as the ac-Stark shift \cite{Blais2004}, dependent on the average resonator photon number $n_{\rm{r}}$ \cite{Schuster2005}. As expected, the qubit frequency $\tilde{\omega}_{\rm{q}}=\omega_{\rm{q}}+(1+2n_{\rm{r}})g^2/\Delta_{\rm{r,q}}$, measured in two-tone spectroscopy with low spectroscopy power ($P_{\rm{s}} \rightarrow 0$), depends linearly on the resonator drive power $P_{\rm{p}}$ [Fig.~\ref{fig:ResDQDChar}(b)]. Using the independently determined coupling constant $g$ and detuning $\Delta_{\rm{r,q}}$ [Fig.~\ref{fig:ResDQDChar}(a)], we calibrate the average photon number $n_{\rm{r}}$ in the resonator \textsl{vs.}~input power $P_{\rm{p}}$ from the observed linear shift of the qubit frequency [Fig.~\ref{fig:ResDQDChar}(b)] \cite{Schuster2005}.

\begin{figure}[!b]
\includegraphics[width=\columnwidth]{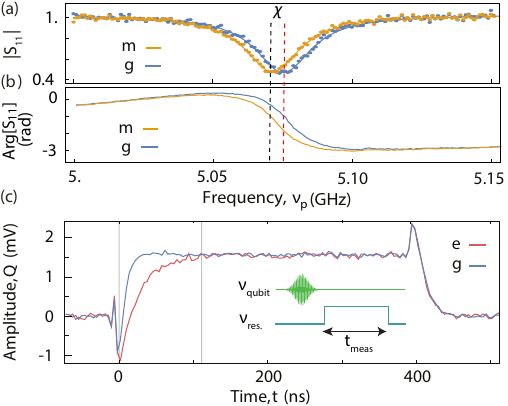}
\caption{Resonator response \textsl{vs.}~measurement frequency $\nu_{\rm{p}}$ and time. (a) Amplitude and (b) phase of the resonator reflectance with no qubit drive tone applied (blue) ($g$) and with a continues tone (orange) at $\nu_\textrm{q}$ resulting in a mixed state (m). $\nu_{\rm{q}}= 5.68\,\rm{GHz}$, $g/2\pi\sim55\,\rm{MHz}$ and $\chi/2\pi \sim 5\,\rm{MHz}$. The vertical dashed red line indicates the resonator read-out frequency selected for the time-resolved  measurements in panel (c). (c) Time-domain response of the Q-quadrature of the resonator reflectance with no pulse ($|g\rangle$, blue curve) and a $\pi$-pulse ($|e\rangle$, red curve) applied at $\nu_\textrm{q}$, respectively.
during the readout stage.
See inset for pulse sequence.
Typical resonator read-out time is $t_\textrm{meas}=400$~ns.
}
\label{fig:Resonator_Response}
\end{figure}

\begin{figure*}[!t]
\includegraphics[width=\textwidth]{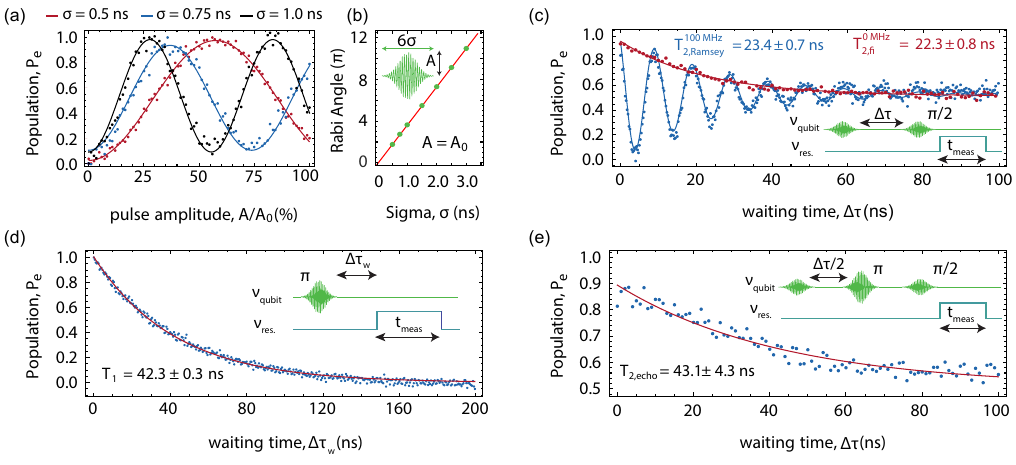}
\caption{Time resolved measurements of the DQD qubit at $\nu_\mathrm{q}(\delta=0)=2t=4.033\,\rm{GHz}$.
(a) Rabi-oscillations for the indicated pulse widths $\it{vs.}$ pulse amplitude $A$ ($A_0 \sim150\,\rm{mV}$). The pulse applied to the qubit is shown in the inset of (b).
(b) Measured Rabi angle (dots) for fixed pulse amplitude $A_0$ \textsl{vs.}~pulse length $\sigma$ with linear fit (red line). (c) Ramsey fringe measurement. The qubit excited state population $P_{\mathrm{e}}$ \textsl{vs.}~time difference $\Delta \tau$ separating $\pi$/2-pulses applied at the qubit frequency $\nu_{\mathrm{q}}$ (red trace) and $100\,\rm{MHz}$ detuned (blue trace).
Pulse sequence in inset. (d) Measurement of DQD qubit relaxation time $T_1$. $\pi$-pulse applied at the qubit frequency $\nu_{\rm{q}}$ followed by a readout pulse with delay $\Delta\tau_{\rm{w}}$.
(e) Spin-echo measurement. See text for details. All data point are averaged 1.6 million times.
}
\label{fig:Time_domain}
\end{figure*}

From the dependence of the qubit linewidth $\delta \nu_q$ on the spectroscopy power ($P_{\rm{s}}$) \cite{Stockklauser2017} at $2t \sim3.3\,\rm{GHz}$ and $\delta=0$, we extract the qubit decoherence rate ${\gamma_2 (P_{\rm{s}} \rightarrow 0)}/{2\pi}={\gamma_1}/{4\pi}+{\gamma_{\phi}}/{2\pi} = (3.3 \pm 0.2)\,\rm{MHz}$ corresponding to $T_2 \sim (48 \pm 2)\,\rm{ns}$. This rate is almost 10 times lower than previously reported values in similar GaAs based devices \cite{Stockklauser2017} and is comparable to decoherence rates found for DQD charge qubits in undoped SiGe heterostructures~\cite{Mi2017}.

To prepare for the time-resolved readout of the DQD qubit, we determine the state-dependent cavity frequency shift $\chi=g^2/\Delta_{\rm{r,q}}$. First, we extract the resonator frequency from a measurement of the reflection coefficient when leaving the qubit in the ground state $|g\rangle$ [blue trace in Figs.~\ref{fig:Resonator_Response}(a,b)]. Then, we apply a continuous coherent tone of duration $t_{\rm{dr}}\gg T_2$ at frequency $\nu_{\rm{s}}=\nu_{\rm{q}}(\delta=0)$ saturating the qubit transition and creating a fully mixed qubit state ($P_{\rm{g}}=P_{\rm{e}}=1/2$). Comparing the frequency of the resonator with the qubit in the fully mixed state [orange curve in Figs.~\ref{fig:Resonator_Response}(a,b)] to the one in the ground state $|g\rangle$, we extract the resonator frequency shift $\chi/2\pi \sim 5\,\rm{MHz}$ \cite{Blais2004}.

By applying resonant microwave pulses of controlled amplitude and duration, we coherently control the quantum state of the DQD charge qubit. We subsequently infer the qubit state by applying a microwave pulse to the resonator and measure its amplitude and phase response using the pulse scheme shown in the inset of Fig.~\ref{fig:Resonator_Response}(c). This approach was previously demonstrated for superconducting qubits \cite{Wallraff2005,Bianchetti2009}.

We record the time-dependent resonator response to the applied measurement microwave pulse with the qubit left in the ground state $|g\rangle$ (blue trace) and when applying a microwave pulse to the qubit preparing the excited state $|e\rangle$ (red trace) [Fig.~\ref{fig:Resonator_Response}(c)]. We adjust the phase of the measurement pulse to maximize (minimize) the detected signal in the Q (I) quadrature. All measurements are performed at $\delta=0$ to maximize qubit coherence.

When applying the readout pulse with the qubit in $|g\rangle$, we observe an exponential rise of the resonator response reaching a steady state on a time scale of $\sim 1/\kappa$. In $|e\rangle$, the resonator frequency is shifted by $2\chi$ resulting in a different Q quadrature response (red trace).
The integrated area between the two curves in Fig.~\ref{fig:Resonator_Response}(c) is proportional to the qubit excited state population $P_{\rm{e}}$ as discussed in detail, for example, in Refs.~\onlinecite{Bianchetti2009,Wallraff2005}.

We apply microwave drive pulses to the DQD qubit at its transition frequency $\nu_{\rm{dr}}=\nu_{\rm{q}}(\delta=0)=2t=4.033\,\rm{GHz}$ through the resonator. The qubit is detuned by $\Delta_{\rm{r,q}}\sim 15\, g\gg \kappa_{\rm{tot}}$ from the resonator. We synthesize the Gaussian qubit control pulses directly (without upconversion) by using an arbitrary waveform generator (AWG) with 25 GS/s sampling rate allowing for good pulse definition down to sub-ns pulse length.

We observe Rabi oscillations in the DQD charge qubit excited state population $P_{\rm{e}}$ by applying pulses [see inset of Fig.~\ref{fig:Time_domain}(b)] with standard deviation $\sigma \sim 0.5,\,0.75$ and $1.0 \, \rm{ns}$ \textsl{vs.}~normalized microwave pulse amplitudes $A/A_0$ followed by pulsed dispersive readout as described above [Fig.~\ref{fig:Time_domain}(a)].
The fastest $\pi$-Rabi pulse, realized in our experiment by using the maximum available pulse amplitude $A_0 \sim150\,\rm{mV}$ at the AWG output, has a standard deviation of $\sigma \sim0.25\,\rm{ns}$. This corresponds to a sizable Rabi frequency of up to $\sim 800 \,\rm{MHz}$ averaged over the pulse duration.
We observe the expected linear dependence of the Rabi angle \textsl{vs.} pulse length obtained for a fixed maximum pulse amplitude [Fig.~\ref{fig:Time_domain}(b)].

We determine the coherence time of the DQD charge qubit, configured as in Fig.~\ref{fig:Time_domain}(a), from a Ramsey fringe experiment using two $\pi/2$ pulses separated by a free evolution time $\Delta \tau$ followed by a readout pulse [inset, Fig.~\ref{fig:Time_domain}(c)]. Driving the qubit on resonance $\Delta_{\rm{q,dr}}= \omega_{\rm{q}}-\omega_{\rm{dr}}=0$ (red curve) or $\Delta_{\rm{q,dr}}= \omega_{\rm{q}}-\omega_{\rm{dr}}=2\pi100\,\rm{MHz}$ detuned (blue curve),
we obtain a free induction decay time of $T_{2,\rm{fi}} \sim (22.3\pm 0.8)\,\mathrm{ns}$ or a Ramsey decay time of $T_{2,\rm{Ramsey}} \sim (23.4\pm 0.7)\,\mathrm{ns}$ when extracting the exponential decay coefficient from the data [Fig.~\ref{fig:Time_domain}(c)].

We determine the energy relaxation time $T_1\sim (42.3 \pm 0.3)\,\mathrm{ns}$ of the DQD charge qubit in the same configuration by first initializing the qubit in $|e\rangle$ and varying the time $\Delta \tau_w$ before reading out the qubit state [Fig.~\ref{fig:Time_domain}(d)]. In this specific DQD configuration $T_2 \ll 2T_1$, indicating that coherence is limited by pure dephasing.
To investigate the origin of the low frequency noise limiting coherence, we also perform a Hahn echo experiment by interleaving the Ramsey sequence with an extra $\pi$-pulse in the middle [inset, Fig.~\ref{fig:Time_domain}(e)]. The echo decay time $T_{2,\rm{echo}}\sim (43.1\pm 4.3) \,\mathrm{ns}$ [Fig.~\ref{fig:Time_domain}(e)] is a factor of 2 longer than the $T_{2,\rm{Ramsey}}$ but still lower than $2T_1$, indicating that fluctuations faster than $\Delta \tau$ contribute to dephasing.

Dispersive read-out combined with all-microwave control of qubits is an essential feature of quantum information processing with superconducting circuits. This work demonstrates that these assets can also come to fruition in circuit QED with semiconductor qubits. We are convinced that the presented methods will contribute significantly to the continued improvement of tools and techniques for quantum information processing with charge and spin qubits in semiconductor nano-structures. In particular, the methods presented here do allow for a detailed study of coherence properties of charge qubits, the results of which we will present elsewhere.

We acknowledge useful discussions with Andreas Landig, Theodore Walter, Philipp Kurpiers, Anton Poto\v{c}nik, Christian K. Andersen and Johannes Heinsoo. We thank Alexandre Blais for valuable feedback on the manuscript. This work was supported by ETH Zurich and in part by the Swiss National Science Foundation through the National Center of Competence in Research (NCCR) Quantum Science and Technology.

\begingroup

\def\refname{References}

\def\bibname{References}

\endgroup


\newpage

\appendix

\onecolumngrid

\begin{center}


\vspace{24pt}

{\bf \Large Supplemental Material}

\end{center}

\vspace{24pt}

\twocolumngrid


\addtocontents{toc}{\setcounter{tocdepth}{0}}

\setcounter{figure}{0}

\renewcommand{\thefigure}{S\arabic{figure}}

\section{Setup Details and Sample Characterization}

In this section, we describe in more detail the measurement setup and the device parameters.

\begin{figure*}[!b]
\includegraphics[width=0.7\textwidth]{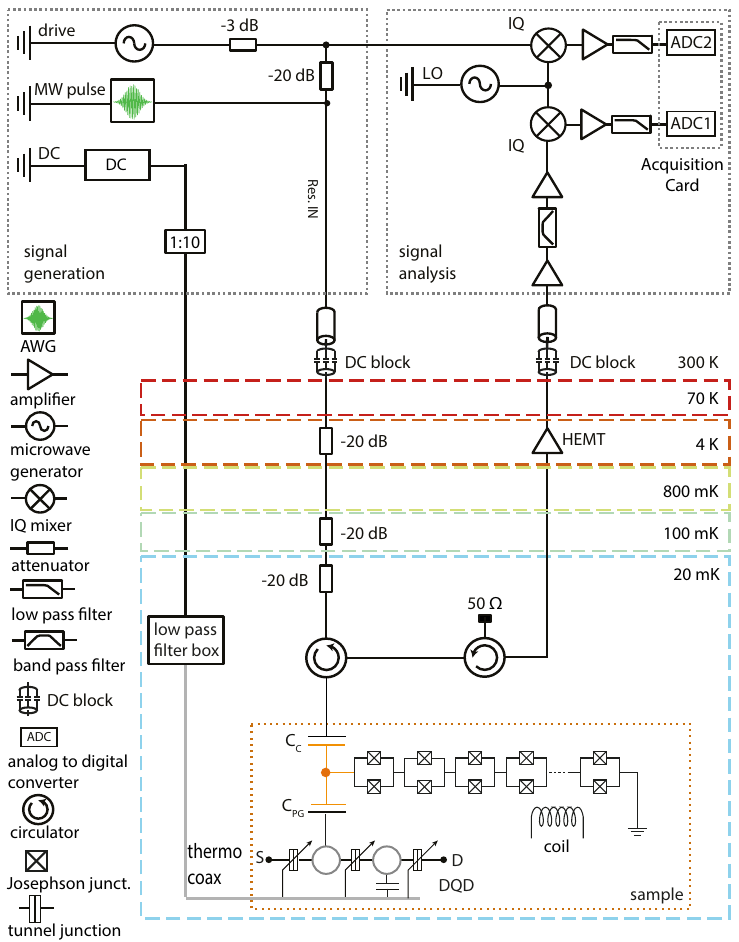}
\caption{
Simplified circuit diagram of the RF setup. See text for details.
}
\label{fig:Setup}
\end{figure*}

The measurement setup consists of DC electronics for biasing and reading out the quantum dot and RF electronics to probe the resonator [Fig.~\ref{fig:Setup}]. The DQD gates are biased using a dc voltage sources (Yokogawa7651). The DC lines contain $1:10$ voltage dividers for noise reduction at room temperature and are low pass filtered at base temperature. This filter is connected to the sample holder by thermocoax lines \cite{Zorin1995}, which provide additional filtering in the microwave range. A symmetric current-to-voltage (IV) converter is used to apply a voltage bias to the 2DEG and measure the resulting current.
The room temperature electronics consists of signal generation and signal analysis components as sketched in Fig.~\ref{fig:Setup}. A microwave generator provides the input signal of the resonator (drive). \\
For the time-resolved measurements, Gaussian pulses with derivative removal via adiabatic gate (DRAG) technique \cite{Chow2010} are generated by an arbitrary wave form generator with 25 GS/s and 10 bit resolution (Tektronix AWG70002) directly at the qubit frequency $\nu_\mathrm{q}$. The DRAG technique is used to minimize the frequency component at the SQUID array resonator frequency $\nu_\mathrm{r}$. 
The resonator drive tone for read-out is gated. \\
The output field is routed through two circulators before reaching a high electron mobility transistor (HEMT) amplifier located on the $4$ K plate of the cryostat. This amplifier has a gain of $39$ dB and a noise temperature of 6 K.
The signal is further amplified and filtered at room temperature and converted to an intermediate frequency (IF) of 250 MHz. In the down-conversion process the signal is mixed with a local oscillator (LO), which is 250 MHz detuned from the signal to be detected [signal analysis box in Fig.~\ref{fig:Setup}]. The same mixing process creates a phase reference signal which is split off the input signal of the resonator (drive) and does not pass through the sample.
The down-converted signals are subsequently amplified, filtered and digitized using an analog-to-digital converter (ADC) outputting its data to a field-programmable gate array (FPGA) signal processing board or to an acquisition card. Digital down-conversion to zero frequency and digital filtering yield the quadrature amplitudes I and Q or equivalently the amplitude A and phase $\phi$ of the complex signal amplitude $S=I+iQ=Ae^{i\phi}$ \cite{Wallraffs2004}.\\

The electrostatics of the DQD is tuned predominantly by making use of the left and right side gates (LSG, RSG). The plunger gates, not dc-biased, are connected to the SQUID array resonator (LPG) and to the groud (RPG). Additionally, the slightly modified DQD gate layout [see Fig.~1(b)] of this device (compared to the gate layout in \cite{Stockklausers2017}) allows to form the QDs closer to each other and increases the lever arm of the plunger gates, directly placed above the desired dot positions. Furthermore, the SQUID array resonator is better shielded from the metallic gates by additional Al ground plane. The coupling capacitance of the SQUID array to the drive line has been increased (with respect to the previous design in \cite{Stockklausers2017}) by extending the metal structure of the drive line around the array island [see Fig.~1(a)].

The extracted device parameters differ significantly from the ones of the previous device generation reported in \cite{Stockklausers2017}. While the external linewidth of the resonator is increased due to the larger coupling capacitance of the SQUID array island to the drive line ($C_{\mathrm{c}} \sim 12\, \mathrm{fF}$ and $\kappa_{\mathrm{ext}}= C_{\mathrm{c}}^2 \omega_{\mathrm{r}}^3 Z_{\mathrm{TL}}Z_{\mathrm{r}}/4$) \cite{Wong2017},
the internal linewidth is reduced.

\section{Charge Qubit Spectroscopy}
\begin{figure}[!b]
\includegraphics[width=\columnwidth]{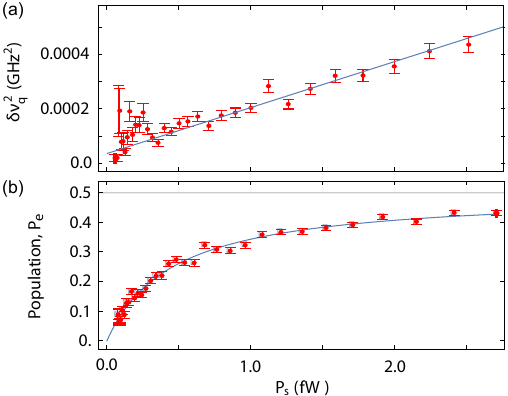}
\caption{
(a) Extracted qubit linewidth $\delta \nu_q$ vs. the power of the spectroscopy tone $P_{\rm{s}}$, evaluated at the DQD sweetspot ($\delta=0$) and $2t\sim 3.3$ GHz. The blue solid line represents a linear fit which allows to extract a $\delta \nu_q(P_{\rm{s}} \rightarrow 0)\sim 3.3 \pm 0.2$ MHz. (b) Saturation of qubit population
with spectroscopy drive power $P_{\rm{s}}$. $P_{\rm{e}}$ is the excited state population of the qubit, which saturates at 1/2.
}
\label{fig:qubitSpectr}
\end{figure}

We further analyze the spectroscopy power ($P_{\mathrm{s}}$) dependence of the qubit linewidth $\delta \nu_{\mathrm{q}}$, reported in Fig.~\ref{fig:qubitSpectr}(a), from which we extract the qubit decoherence rate $\gamma_2 (P_{\mathrm{s}} \rightarrow 0)/2\pi=\gamma_1/4\pi+\gamma_{\phi}/2\pi= (3.3 \pm 0.2)\,\rm{MHz}$ corresponding to $T_2 \sim (48 \pm 2)\,\rm{ns}$, evaluated for $2t=3.29\,\rm{GHz}$ and $\delta=0$ (for the same DQD charge configuration explored in the main text). This value is almost 10 times better than what we previously reported on a similar device in \cite{Stockklausers2017} and comparable to what observed in \cite{Mis2017} for a DQD charge qubit in undoped SiGe heterostructures.
Increasing the drive strength $P_{\mathrm{s}}$ we observe the qubit transition to approach saturation [Fig.~\ref{fig:qubitSpectr}(b)] \cite{Schusters2005,Stockklausers2017}.

%
%
%
%


%

\end{document}